\documentclass[aps,preprint,preprintnumbers,amsmath,amssymb]{revtex4}
\usepackage{amsmath,mathrsfs,amsbsy,graphicx,bm,amsthm,amsfonts}
\usepackage{units}
\usepackage{bbm}
\usepackage{times}
\usepackage{dcolumn}
\usepackage{mathrsfs}% Align table columns on decimal point
\usepackage{amsmath,amssymb,epsfig}
\usepackage{color}
\usepackage{hyperref}
\def\aj{Astron. J.}
\def\aap{Astron. Astrophys.}
\def\mnras{Mon. Not. Roy. Astron. Soc.}
\def\apjl{Astrophys. J. Lett.}
\def\jcap{JCAP}
\def\apjs{Astrophys. J., Suppl. Ser.}
\def\aapr{Astron. Astrophys. Rev.}
\def\pasp{Publications of the Astronomical Society of the Pacific}
\newcommand{\LCDM}{\rm{\Lambda}CDM}
\newcommand{\Mpc}{\mathrm{~km~s^{-1}~Mpc^{-1}}}
\hypersetup{
    colorlinks=true,
    linkcolor=red,
    citecolor=blue,
    urlcolor=magenta,
}

\begin{document}

\title{Measurements of the Hubble constant from combinations of supernovae and radio quasars }
%Multiple measurements on the Hubble constant and exploring its possible evolution }
\author{Tonghua Liu$^{1}$, Xiyan Yang$^{1}$, Zisheng Zhang$^{1}$,  Jieci Wang$^{2}$ \footnote{ jcwang@hunnu.edu.cn}, Marek Biesiada$^{3}$ \footnote{Marek.Biesiada@ncbj.gov.pl}}
\affiliation{1. School of Physics and Optoelectronic, Yangtze University, Jingzhou 434023, China; \\
2. Department of Physics, and Collaborative Innovation Center for Quantum Effects and Applications, Hunan Normal University, Changsha, Hunan 410081, China. \\
3. National Centre for Nuclear Research, Pasteura 7, PL-02-093 Warsaw, Poland}

\baselineskip=0.65 cm

%\vspace*{-0.2cm}
\begin{abstract}
In this letter, we propose an improved cosmological model independent method of determining the value of the Hubble constant $H_0$. The method uses unanchored luminosity distances $H_0d_L(z)$ from SN Ia Pantheon data combined with angular diameter distances $d_A(z)$ from a sample of intermediate luminosity radio quasars calibrated as standard rulers. The distance duality relation between $d_L(z)$ and  $d_A(z)$, which is robust and independent of any cosmological model, allows to disentangle $H_0$ from such combination. However, the number of redshift matched quasars and SN Ia pairs is small (37 data-points). Hence, we take an advantage from the Artificial Neural Network (ANN) method to recover the $d_A(z)$ relation from a network trained on full 120 radio quasar sample. In this case, the result is unambiguously consistent with values of $H_0$ obtained from local probes by SH0ES and H0LiCOW collaborations. Three statistical summary measures: weighted mean $\widetilde{H}_0=73.51(\pm0.67) \Mpc$, median $Med(H_0)=74.71(\pm4.08) \Mpc$ and MCMC simulated posterior distribution $H_0=73.52^{+0.66}_{-0.68} \Mpc$ are fully consistent with each other and the precision reached $1\%$ level. This is encouraging for the future applications of our method.
Because individual measurements of $H_0$ are related to different redshifts spanning the range $z=0.5 - 2.0$, we take advantage of this fact to check if there is any noticeable trend in $H_0$ measurements with redshift of objects used for this purpose. However, our result is that the data we used strongly support the lack of such systematic effects.
\end{abstract}

%\vspace*{0.2cm}

\maketitle
\section{Introduction}
Over the last decades, one of the most important achievements in observational cosmology were very precise  measurements of tiny anisotropies in the cosmic microwave background radiation (CMB) \cite{2003ApJS..148....1B,2011ApJS..192...18K,2016A&A...594A..13P}. During the mission of \emph{Planck} satellite not only temperature fluctuations (T) but also E-modes of CMB polarization pattern have been accurately measured and TT, TE, EE power spectra have been precisely determined up to very high multipole moments $l\sim 2500$ (see \citep{Aghanim2020} and references therein). The power spectrum of temperature fluctuations revealed the so called acoustic peaks enabling direct, empirical studies of the early Universe and initial conditions for the formation of the large scale structure. CMB data combined with baryonic acoustic oscillations (BAO) measurements (\cite{2020NatAs...4..196D,BAO,BAO1,BAO2,BAO3,BAO4,BAO5} and references therein) in large galaxy catalogs led modern cosmology into an era of precision cosmology. However, with increased precision accuracy issues emerged \cite{2022JHEAp..34...49A,2021APh...13102605D,2022NewAR..9501659P}. One of the most widely known issue is the inconsistency between the values of the Hubble constant $H_0$ (current expansion rate of the Universe) obtained by different techniques. The CMB measurement using the \emph{Planck} data yielded $H_0=67.4 \pm 0.5 \Mpc$ at the $68\%$ confidence level (CL) \citep{Aghanim2020}. This result is in tension of more than $4\sigma$ with the value of $H_0=73.2 \pm 1.3 \Mpc$
at the $68\%$ CL reported by SH0ES (\textit{Supernova $H_0$ for the Equation of State}) collaboration  using Type Ia supernovae (SN Ia) calibrated by local Cepheid variable stars \citep{SH0ES}. Such tension has reached 4.4$\sigma$-6$\sigma$ with the accumulation of precise astrophysical observations. It should be emphasized that the Hubble constant obtained from the CMB data requires the assumption of $\Lambda$CDM cosmological model.

The discrepancy between just two methods necessitates involvement of independent alternative techniques. Excellent and comprehensive review by \citep{DiValentino:2021} contains a detailed discussion of such alternative methods and current results obtained within their frameworks. Two of them are worth mentioning here. First is the possibility to directly measure the Hubble constant from time-delays between multiple images in strongly lensed systems \citep{TreuMarshall}. The H0LiCOW ($H_0$ Lenses in COSMOGRAIL's Wellspring) collaboration reported the value of the Hubble constant $H_0=73.3^{+1.7}_{-1.8} \Mpc$ obtained from the joint analysis of time-delay measurements of six lensed quasars with the assumption of a spatially flat $\LCDM$ model \citep{Wong:2019kwg}. See refs \citep{2014ApJ...788L..35S,2019A&A...628L...7T,2018MNRAS.481.1041T,Liu12} for more on the strong lensing time-delay method to measure $H_0$. Another alternative method comes from the standard sirens, which is promising in the era of gravitational wave (GW) astronomy \cite{2009LRR....12....2S,2015JPhCS.610a2021V,2017ApJ...848L..12A}. Namely, from the detected waveform of coalescing binary systems one is able to measure the so called chirp mass and luminosity distance \cite{1986Natur.323..310S,2016PhRvL.116f1102A,2017PhRvL.119p1101A}. Besides the parallax method, this is the only one possibility in astronomy and cosmology where distance can be directly measured without reference to the distance ladder calibration. The major limitation is that in GW domain redshift of the source is rarely available. The famous event GW170817 whose optical counterpart has been identified provided such measurement \citep{GW170817a} yielding $H_0=70.0^{+12.0}_{-8.0} \Mpc$, which after further correction for peculiar motion has been improved \citep{GW170817b} to $H_0=68.3^{+4.6}_{-4.5} \Mpc$. These results are still far from the precision cosmology standards.
Since there is no evidence of considerable systematic uncertainties in either the \emph{Planck} data \citep{DiValentino:2021,Aghanim2020} and the local measurements \citep{2022ApJ...938..111B,2019ApJ...876...85R,Follin:2017ljs}, one natural way is to seeks the source of this  discrepancy in modifications of the cosmological model. Thus, there is a growing interest in  alternative cosmological models beyond $\LCDM$, such as the Early Dark Energy models \citep{2016PhRvD..94j3523K,2023arXiv230300388T,2022arXiv221104492K},  interacting dark energy models \citep{2010MNRAS.402.2355V,2017JCAP...10..030Z,2016RPPh...79i6901W,2012AIPC.1471...51Z},  modified gravity $f(R)$ \citep{2010RvMP...82..451S,2010LRR....13....3D,2023arXiv230406425L} and $f(T)$ \citep{2011PhRvD..84b4020H,2016RPPh...79j6901C} models, to give just a few examples. All this  highlights the importance of any alternative, cosmological model independent methods of measuring the Hubble constant.

Recently, a cosmological model-independent approach to assess the Hubble constant was proposed by \citet{2023PhRvD.107b3520R}.  Their method relies on the distance duality relation (DDR) \citep{1933PMag...15..761E},
which links together luminosity distance $d_L(z)$ and angular diameter distance $d_A(z)$. The DDR is robust, i.e. valid in any metric theory of gravity (not only in Friedmann-Lema{\^i}tre-Robertson-Walker (FLRW) model) provided that number of photons in the beam is conserved (i.e. negligible absorption). For this purpose they used un-anchored SN Ia as source of $d_L(z)$, line of sight and transverse BAO data as a source of $H(z)d_A(z)$ and combination and cosmic chronometers as a source of $H(z)$. Technically, the distance reconstruction steps necessary in their methodology were performed using the Gaussian process (GP) regression. The bottleneck of their study is the scarcity of $d_A(z)$ data points: 7 BAO data points \cite{BAO3,2021MNRAS.500..736B,2020MNRAS.498.2492G,2020MNRAS.499.5527T,2021MNRAS.501.5616D,2020MNRAS.499..210N,2021MNRAS.500.1201H,2019A&A...629A..85D,2019A&A...629A..86B} and 30 data points for cosmic chronometers \cite{2014RAA....14.1221Z,2010JCAP...02..008S,2012JCAP...08..006M,2016JCAP...05..014M,2015MNRAS.450L..16M}.

In this letter we extend the work of \citet{2023PhRvD.107b3520R} in two important aspects. First, we use the sample of 120 intermediate-luminosity radio quasars compiled in \citet{2017A&A...606A..15C} as a source of $d_A(z)$. Besides much richer sample, this allows us to follow the methodology of \citet{2023PhRvD.107b3520R} directly, without need to use Hubble functions $H(z)$ from cosmic chronometers. Second improvement is that instead of Gaussian processes we use a machine learning method -- Artificial Neural Network (ANN) algorithm to perform necessary reconstruction steps. More detailed description of the methodology, the samples used and the ANN method will be given below in Section II. In Section III, we show our results and discussion. We conclude in Section IV.

\section{Methodology and observational data}
\subsection{Methodology of measuring the Hubble constant}
Modern cosmology is based on the notion of homogeneous and isotropic Universe, well supported by the high degree of CMB isotropy. The geometry of space-time in the largest scales is described by the FLRW metric
\begin{equation}
ds^2=c dt^2-\frac{a(t)^2}{1-Kr^2}dr^2-a(t)^2r^2d\Omega(\theta,\phi)^2,
\end{equation}
where $c$ is the speed of light, $a(t)$ is the scale factor, and $K$ is dimensionless curvature taking one of three values $\{-1, 0, 1\}$ corresponding to closed, flat and open universe, respectively. The cosmic curvature parameter $\rm\Omega_K$ is related to $K$ and the Hubble constant $H_0$, as $\rm{\Omega_K}$ $=-c^2K/a^2_0H_0^2$.
It is evident, from the FLRW metric, that spatial lengths (e.g. wavelengths or distances between galaxies) change in time according to the scale factor $a(t)$. The scale factor is therefore related to the redshift $z$ of the source by $a=1/(1+z)$ and the expansion rate at redshift $z$ is $H(z) = {\dot a}/a$.
In the curved FLRW space-time the issue of distances becomes subtle. Namely, the distance to the object at redshift $z$ suggested by the FLRW metric is the comoving distance $d_C(z)$ given by the relation
\begin{equation}
d_C(z) = c\int^z_0\frac{dz'}{H(z')}.
\end{equation}
Unfortunately the comoving distance is unobservable. Two closely related distances, which can be measured are the luminosity distance $d_L(z) = d_C(z) (1+z)$ and the angular diameter distance $d_A(z) = d_C(z) / (1+z)$. The luminosity distance is usually inferred from sources with known intrinsic brightness or standardized luminosity, and the angular diameter distance is derived from the angular scale of objects whose intrinsic sizes are known.
If the number of photons traveling along the null geodesics between the observer and the source is conserved, then $d_L(z)$ and $d_A(z)$ should satisfy the following relation %at the same redshift
\citep{1933PMag...15..761E}, %2007GReGr..39.1055E},
\begin{equation} \label{DDR}
d_L(z)=d_A(z)(1+z)^2.
\end{equation}
This relation is also known as the DDR and is valid in any metric theory of gravity.
Therefore, the DDR is completely independent of the cosmological model. Up to now, numerous works \citep{2021ApJ...909..118Z,2023PhLB..838137687,2020ApJ...892..103Z,2021PhRvD.103j3513A,2019ApJ...885...70L} focused on  testing the DDR with the real data, demonstrated lack of evidence for noticeable departure from the DDR. According to the philosophy of \citet{2023PhRvD.107b3520R} one can rewrite Eq.(\ref{DDR}) in the form
\begin{equation} \label{H0}
H_0=\frac{1}{(1+z)^2}\frac{H_0d_L(z)}{d_A(z)}.
\end{equation}
It is now clear that in order to obtain $H_0$ one has to measure the unanchored luminosity distance $H_0 d_L(z)$ and angular diameter distance at the same redshift $z$. As it was already mentioned, we take advantage of SN Ia as standard candles and intermediate luminosity radio quasars as standard rulers. In particular the redshift range of these two samples is similar. Below, we will present these samples in more details.

\subsection{Unanchored luminosity distance from SN Ia}

SN Ia observations led to the discovery the accelerating expansion of the Universe \cite{Riess1998,Perlmutter1999}. They are incisive probes of cosmology through determining the shape of the Hubble diagram, i.e. the luminosity distance vs. redshift relation.
However, the absolute distance $d_L(z)$ is entangled with the combination of the absolute magnitude of SN Ia and the Hubble constant. Therefore, observations of SN Ia can directly provide the unanchored luminosity distance $H_0d_L(z)$, which is exactly the quantity we need.
The unanchored luminosity distance can be derived from apparent magnitude
\begin{equation} \label{unanchored}
m_B=5\log_{10}(H_0d_L(z))-5a_B,
\end{equation}
where we adopt the value $a_B=0.71273\pm0.00176$ as inferred in work \citep{2016ApJ...826...56R}.
%from the same SN Ia measurements that we are using.
In a recent paper,  \citet{2022ApJ...934L...7R} proposed an improved treatment of the intercept $a_B$ valid for an arbitrary expansion history in terms of deceleration parameter $q_0$ and jerk $j_0$. We will assume its fixed value as mentioned above.

Over the past two decades, many supernova surveys have focused on detecting supernovae within a  considerable range of redshifts, including low redshifts ($0.01<z<0.1$), e.g. CfA1-CfA4, CSP and LOSS \citep{Riess99,Jha06,Stritzinger11}, and four main surveys probing the $z>0.1$ redshift range like ESSENCE, SNLS, SDSS and PS1 \citep{Miknaitis07,Conley11,Frieman08,2014ApJ...795...45S}. More high redshift observation of supernovae, like SCP, GOODS and CANDELS/CLASH surveys released the high-z $(z > 1.0)$ data \citep{Suzuki12,Riess04,Riess07,Rodney14}. More recently, \citet{2018ApJ...859..101S} combined the subset of 279 Pan-STARRS1(PS1) ($0.03 < z < 0.68$) supernovae \citep{2014ApJ...795...44R,2014ApJ...795...45S} with  useful data of SN Ia from SDSS, SNLS, and various low redshift and HST samples to form the largest combined sample of SN Ia consisting of a total of 1048 SNe Ia ranging from $0.01<z<2.3$, which is known as the  ``Pantheon Sample". We refer to work \citep{2018ApJ...859..101S} for more details about the SN Ia standardization process including the improvements of the PS1 SNe photometry, astrometry and calibration.
In order to obtain unanchored luminosity distances given by Eq.(\ref{unanchored}), we use the Pantheon dataset. The scatter diagram of the apparent magnitude for 1048 observed SN Ia is shown in the left panel of Fig.~1.

\begin{figure}
\centering
\includegraphics[width=8.1cm,height=6cm]{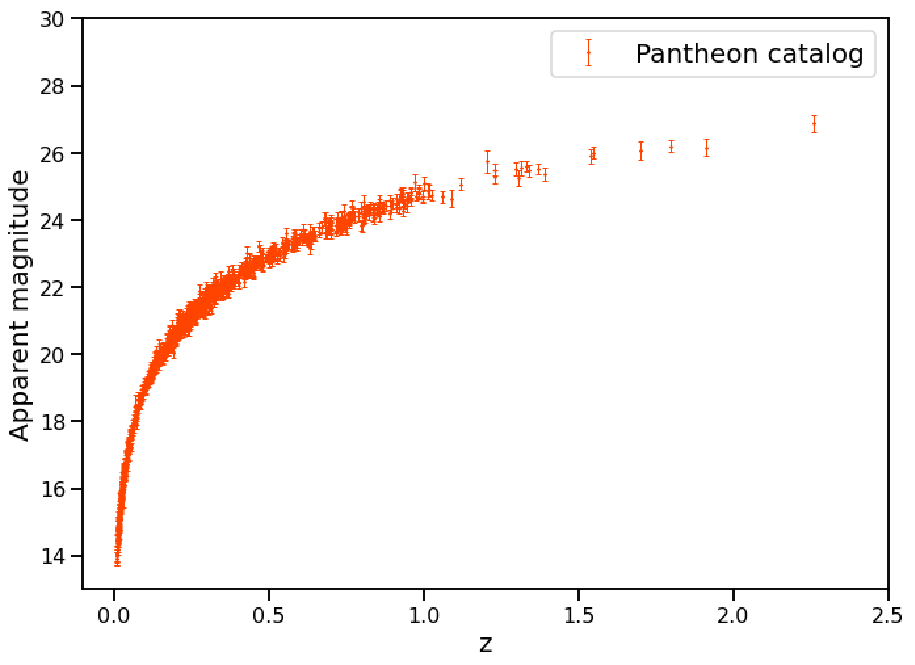}
\includegraphics[width=8.2cm,height=6cm]{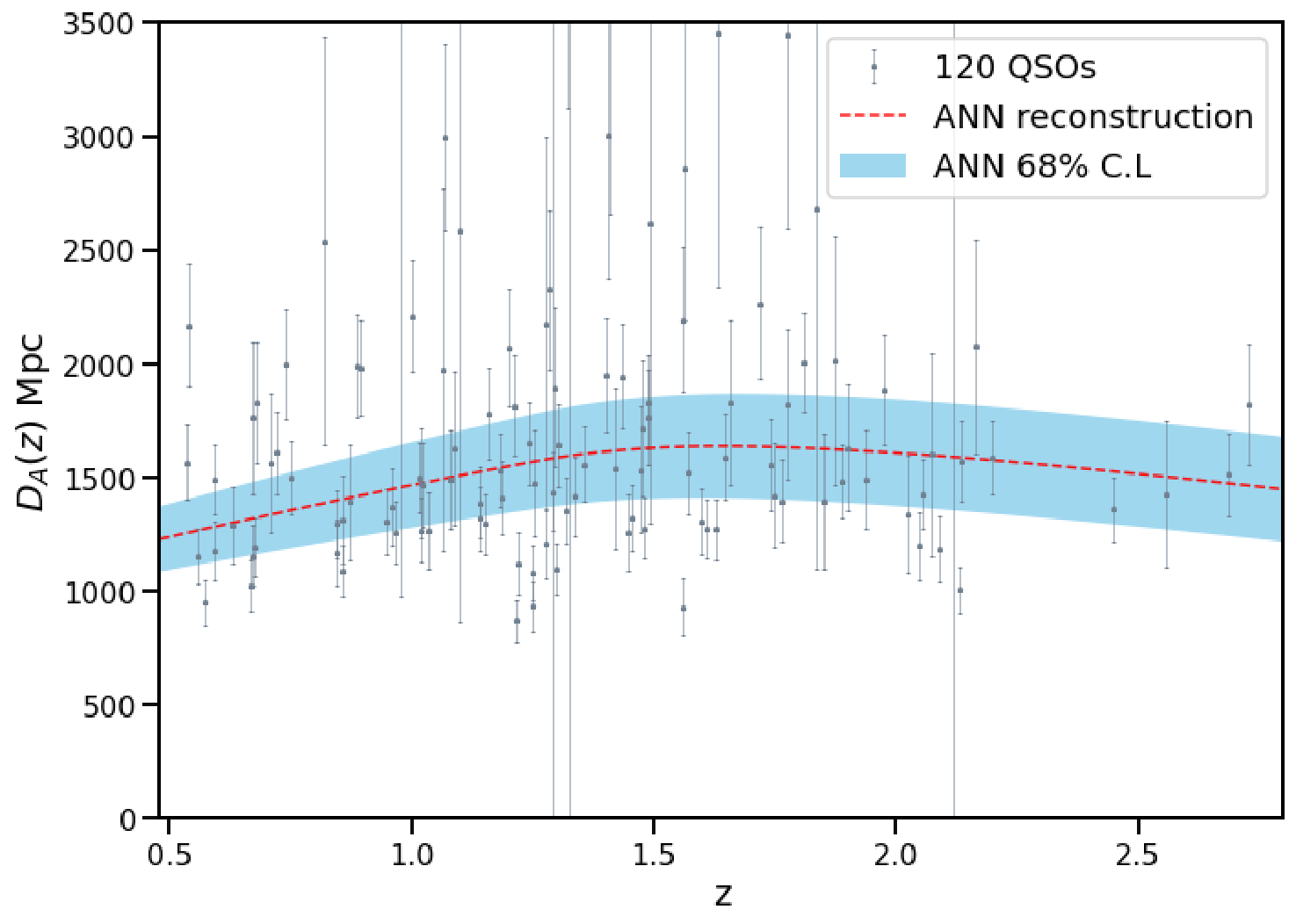}
\caption{\textit{Left panel}: The apparent magnitude vs. redshift for the Pantheon sample of 1048 SN Ia. \textit{Right panel}: Angular diameter distance vs. redshift based on the sample of 120 intermediate luminosity radio quasars. Gray dots with uncertainty bars represent angular diameter distances obtained from observed angular sizes of radio quasars. The ANN reconstructed function $d_A(z)$ is shown as red dotted line and corresponding $1\sigma$ uncertainty band is shown as a blue region.
}
\end{figure}

\subsection{Angular diameter distances from angular size of the compact structure in radio quasars}

The angular size-distance relation in compact radio quasar for cosmological inference was first proposed by \citet{Kellermann93}, who tried to obtain the deceleration parameter with 79 compact radio sources observed by VLBI at 5 GHz. Thereafter, \citet{1994ApJ...425..442G} extended this method and attempted
to investigate the dependence of characteristic size on luminosity and redshift based on 337 Active Galactic Nuclei (AGNs) observed at 2.29 GHz \citep{1985AJ.....90.1599P}. In the subsequent analysis,  \citep{1994ApJ...425..442G} adopted the visibility modulus $\Gamma=S_c/S_t$ to redefine angular size of radio sources $\theta$,  which can be expressed by $\theta={2\sqrt{-\ln\Gamma \ln 2} \over \pi B}$,
where $B$ is the interferometer baseline measured in multiple of wavelengths, $S_c$ and $S_t$
are correlated flux density and total flux density, respectively. Based on a simple geometric relation between the angular size and distance, the angular diameter distance $d_A(z)$ can be written as
\begin{equation}
d_A(z)=\frac{lL^{\beta}(1+z)^n}{\theta(z)},
\end{equation}
where $L$ is the intrinsic luminosity of the source, $\beta$ and $n$ are used to quantify the possible  ``angular size-redshift'' and ``angular size-luminosity'' relations, respectively. The parameter $l$ represents the linear size scaling factor describing the apparent distribution of radio brightness within the core and $\theta(z)$ is the observed angular size measured by VLBI technique.

Our research is based on the sub-sample identified and calibrated in \cite{2017A&A...606A..15C,Cao15,2017JCAP...02..012C}, so a brief description of this sample is appropriate. The original source of the data was a well-known 2.29 GHz VLBI survey undertaken by \citet{1985AJ.....90.1599P} (hereafter called P85). By using a worldwide array of antennas to form an interferometric system %with an effective baseline wavelength,
this survey successfully detected interference fringes from 917 radio sources out of 1398 candidates selected mainly from the Parkes survey \cite{Bolton99}. Subsequently, \citet{Jackson06} updated the P85 sample with respect to redshift, to include a total of 613 objects with redshifts $0.0035\leq  z \leq3.787$. The full listing is available in electronic form at [http://nrl.northumbria.ac.uk/13109/], including source coordinates, redshift, angular
size, uncertainty in the latter, and total flux density. In the subsequent analysis, \citet{2017A&A...606A..15C,2017JCAP...02..012C} divided the Jackson's sample into different sub-samples, according to their optical counterparts and luminosity: low, intermediate, and high-luminosity quasars. It was found that 120 quasars with intermediate-luminosities (ILQSO) have reliable measurements of the angular size of the compact structure from updated 2.29 GHz VLBI survey  with flat spectrum index ($-0.38<\alpha<0.18$). Moreover, they demonstrated that the linear size scaling factor showed negligible dependence on both redshift and intrinsic luminosity ($|n|\simeq10^{-3}, |\beta|\simeq10^{-4}$) \citep{2017JCAP...02..012C,2017A&A...606A..15C}. The sample of 120 intermediate luminosity radio quasars selected in \citep{2017A&A...606A..15C} has been extensively used in various cosmological studies
\citep{2018EPJC...78..749C,2017EPJC...77..891M,2017EPJC...77..502Q,2022A&A...668A..51L}.
The crucial question is about the value of $l$. Similar to absolute magnitudes of SN Ia applied to cosmology, the linear size scaling factor $l$ parameter should also be optimized along with the cosmological model parameters.
The original calibration performed in  \citep{2019PDU....24..274C} through a cosmology-independent Gaussian Process (GP) technique combined with the BAO data resulted with the value of intrinsic linear size $l=11.04\pm0.40$ pc. Hence, in the current analysis, we use this value. The scatter diagram of angular diameter distances obtained from observed angular sizes of 120 intermediate luminosity radio quasars is shown in the right panel of Fig.~1.

\subsection{Measuring $H_0$ with recent observations}
As we describe in more details in the next section, we assessed $H_0$ using two approaches. First, by directly using redshift matched pairs QSO -- SN Ia and second by using ANN reconstructed $d_A(z)$ from radio quasars. In the former case the Eq.(\ref{H0}) can be rewritten in terms of observable quantities in the following way:
\begin{equation} \label{H0_new}
H_0=\frac{\theta10^{0.2(m_B+5a_B)}}{l(1+z)^2},
\end{equation}
while in the latter Eq.(\ref{H0}) is appropriate.
The reason why we use the two methods is that, although the SN Ia and intermediate luminosity radio quasars cover similar redshift range, the number of matched pairs is small (due to relatively small QSO sample size and nonuniform coverage of the redshift range in both samples). Thus we also use the Artificial Neural Network (ANN) method trained on QSO sample to reconstruct $d_A(z)$ function.
The ANN method is a non-parametric approach, which unlike the Gaussian processes does not assume random variables that satisfy the Gaussian distribution. This is a totally data driven approach. The ANN method has been applied to many research fields in astronomy, and has shown excellent performance in cosmological applications %and parameter constraints
\citep{2021PhRvD.103j3513A,2020A&A...644A..80M,2021MNRAS.501.5714W,2020ApJS..249...25W,2022ApJ...936...21Z}. We refer the reader to \citet{2020ApJS..249...25W} for more details about ANN reconstructing $H(z)$ function form cosmic chronometers data, where the  performance of the method was amazing despite the small number of 31 data points. \citet{2020ApJS..246...13W} released the ANN code, and Python module called Reconstruct Functions with ANN (ReFANN) \footnote{https://github.com/Guo-Jian-Wang/refann}. We performed the $d_A(z)$ reconstruction using the 120 intermediate luminosity radio quasars sample.
The final result is shown in the right panel of the Fig. 1.  One can see that the uncertainty band of the ANN reconstructed $d_A(z)$ function is in stark contrast with individual uncertainties and scatter present in QSO data.
%Therefore, the $1\sigma$ confidence region reconstructed by ANN can be considered as the average level of observational error. We refer the reader to refs \citep{2020ApJS..249...25W,2020ApJS..246...13W} for further details on this issue.

\section{Results and discussions}

It is clear from the methodology outline that SN Ia and radio quasars have exactly the same redshift with idealized setting. This does not happen in real data regarding two distinct samples of different objects. One can, however match the objects in narrow redshift bins. In our study we take the matching criterion: $|z_{SN}-z_{QSO}|<0.005$. Such criterion has also been used in works \citep{2011ApJ...729L..14L,2016ApJ...822...74L}. The matching criterion is very restrictive, hence from the samples of 1048 SN Ia and 120 quasars, we are able to select only 37 pairs matched by redshift.
Based on these 37 matched pairs, we calculate $H_0$ according to Eq.(\ref{H0_new}). Uncertainty of individual $H_0$ values are calculated from the standard uncertainty propagation formula, based on the uncorrelated uncertainties of observable quantities including the observed angular size uncertainty $\sigma_{\theta}$, apparent magnitude uncertainty $\sigma_{m_{B}}$, as well as additional systematic uncertainties from $\sigma_{a_B}$ in SN Ia and  $\sigma_{l}$ in radio quasars. The results are shown in Fig.~2. In the left panel of Fig.~2 individual measurements of $H_0$ are shown. As one can see, they display a considerable scatter and diversity of uncertainties. Let us note that this plot also demonstrates that we are able to assess $H_0$ using probes at different redshifts. This circumstance is important in our further discussion. The right panel of Fig.~2 shows the histogram of $H_0$ values obtained from the sample of 37 matched pairs.

\begin{figure}
\centering
{\includegraphics[width=8cm,height=6.3cm]{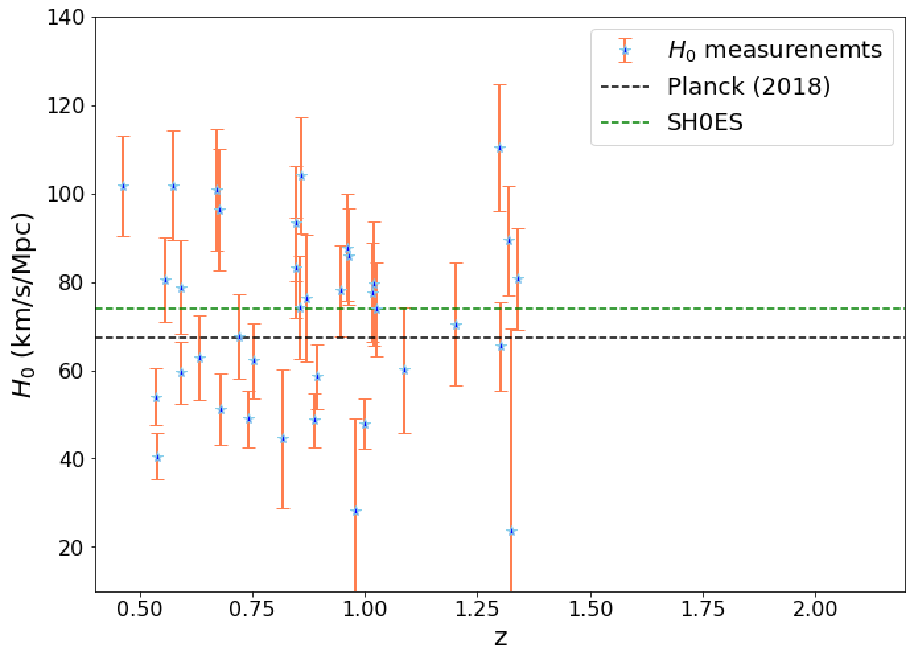}
\includegraphics[width=8cm,height=6.2cm]{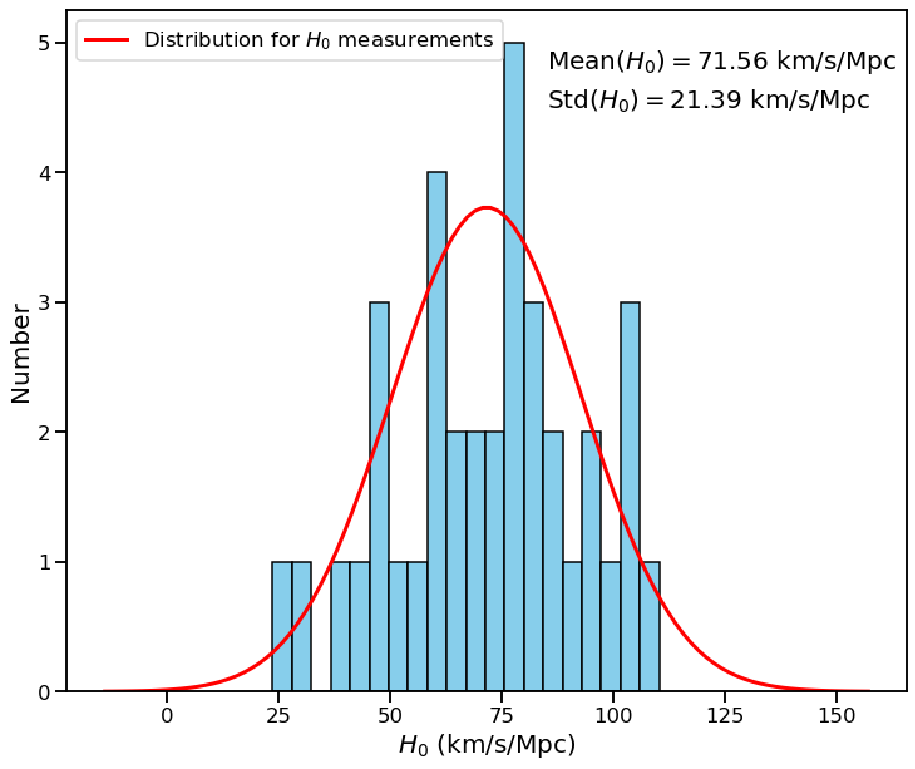}}
\caption{\textit{Left panel}: Individual measurements of $H_0$ based on the 37 pairs of radio quasars and SN Ia matched by redshift. \textit{Right panel}: The histogram plot for the measurements of $H_0$ with best fit Gaussian distribution overplotted.}
\end{figure}

Let us outline the approach. We adopt regarding summary statistics and take advantage of the complete sample of 120 quasars as standardizable rulers (source of $d_A(z)$). Regarding the latter, we decide to reconstruct the $d_A(z)$ function using ANN trained on the full sample of 120 radio quasars. The trained network is able to forecast angular diameter distances at redshifts corresponding to SN Ia in the Pantheon sample. Concerning statistical summary of our results we take three approaches. First is the weighted mean \citep{1993ComPh...7..415B}, which is the way to summarize the data with non-uniform measurement uncertainties. It is the method most often encountered in the literature, in particular in meta-analysis to integrate the results of individual measurements and widely used in statistical analysis of astronomical data \citep{2020MNRAS.496..708L,2019NatSR...911608C}. In our context the weighted mean of $H_0$ reads:
\begin{equation}
\widetilde{H}_0=\frac{\Sigma_i\big(H_{0,i}/\sigma_{H_{0,i}}^2\big)}{\Sigma_i\big(1/\sigma_{H_{0,i}}^2\big)},\,\,\,\,\,\,\,\,\,
\sigma^2_{\widetilde{H}_0}=\frac{1}{\Sigma_i\big(1/\sigma_{H_{0,i}}^2\big)},
\end{equation}
where $\sigma_{\widetilde{H}_0}$ is the uncertainty of $\widetilde{H}_0$.
It is evident that the weighted mean gives more weight to measurements with low uncertainties.
However, it is meaningful for Gaussian random variables, thus we use also a non-parametric summary: the median $Med(H_0)$. It is the most robust measure of central tendency, insensitive to outliers (extreme values) \citep{2012msma.book.....F,2001ApJ...549....1G}. Appropriate non-parametric dispersion measure is the median absolute deviation $MAD(H_0) = Med(|H_{0,i} - Med(H_0)|)$. The third approach we use is the Markov Chain Monte Carlo (MCMC) simulation of the posterior distribution of $H_0$. The summary statistics, i.e. the mean and credible intervals are directly obtained from the posterior. In this MCMC simulation, we adopt  a uniform distribution for the prior distribution of $H_0\in [55, 100] \Mpc$.
Now, getting back to the 37 matched pairs sample, our assessments for the weighted mean and corresponding uncertainty are $\widetilde{H}_0=66.24(\pm1.64) \Mpc$, which is in perfect agreement with the \emph{Planck} 2018 results.
The histogram plot of the individual measurements is given in right panel of the Fig. 2.  From the histogram one may doubt how close the distribution can be approximated by Gaussian. Therefore we perform the Kolmogorov-Smirnov  test \cite{2011arXiv1106.5598B} for Gaussianity, which results with the p-value of $p=0.944$.
Much more sensitive Lilliefors test \cite{2006math.....12708A} give $p=0.758$. Hence, the use of weighted mean as a summary measure is justified.
Within the non-parametric approach, the median value of $H_0$ and corresponding median absolute deviation is $Med(H_0)=74.21(\pm14.77) \Mpc$. The median value is close to the value inferred from the SH0ES collaboration, but the MAD is large. On one side this is a well known property of MAD: it is robust, but not very precise \citep{2012msma.book.....F}. On the other side it reflects a considerable scatter in the small sample. At last the MCMC simulation of the $H_0$ posterior, performed using Python module \textit{emcee}  \citep{2013PASP..125..306F}, give the result: $H_0=64.83^{+1.48}_{-1.58} \Mpc$, where the credible region covers the range between $16^{th}$ and $84^{th} $ percentile. This value is consistent with the weighted mean assessment.

As already stressed, the sample of 37 matched pairs SNIa - QSO is poor and scatter dominated. Therefore we take advantage of the full sample of 120 radio quasars to train the ANN network and recover the $d_A(z)$ relation supported by quasars. Then one is able to forecast angular diameter distance of SN Ia having redshifts in the range of $z$ covered by quasars. This resulted with a sample of 237 SN Ia fully meet this criterion. Uncertainty of reconstructed $d_A(z)$ is obtained from ANN methodology. In principle, $d_A(z)$ from trained network can be projected to the full Pantheon sample, but we do not considered this. The reason is that, in any type of reconstruction, projections beyond the range of data is questionable and burdened with considerable uncertainties.

\begin{figure}
\centering
{\includegraphics[width=8cm,height=6.3cm]{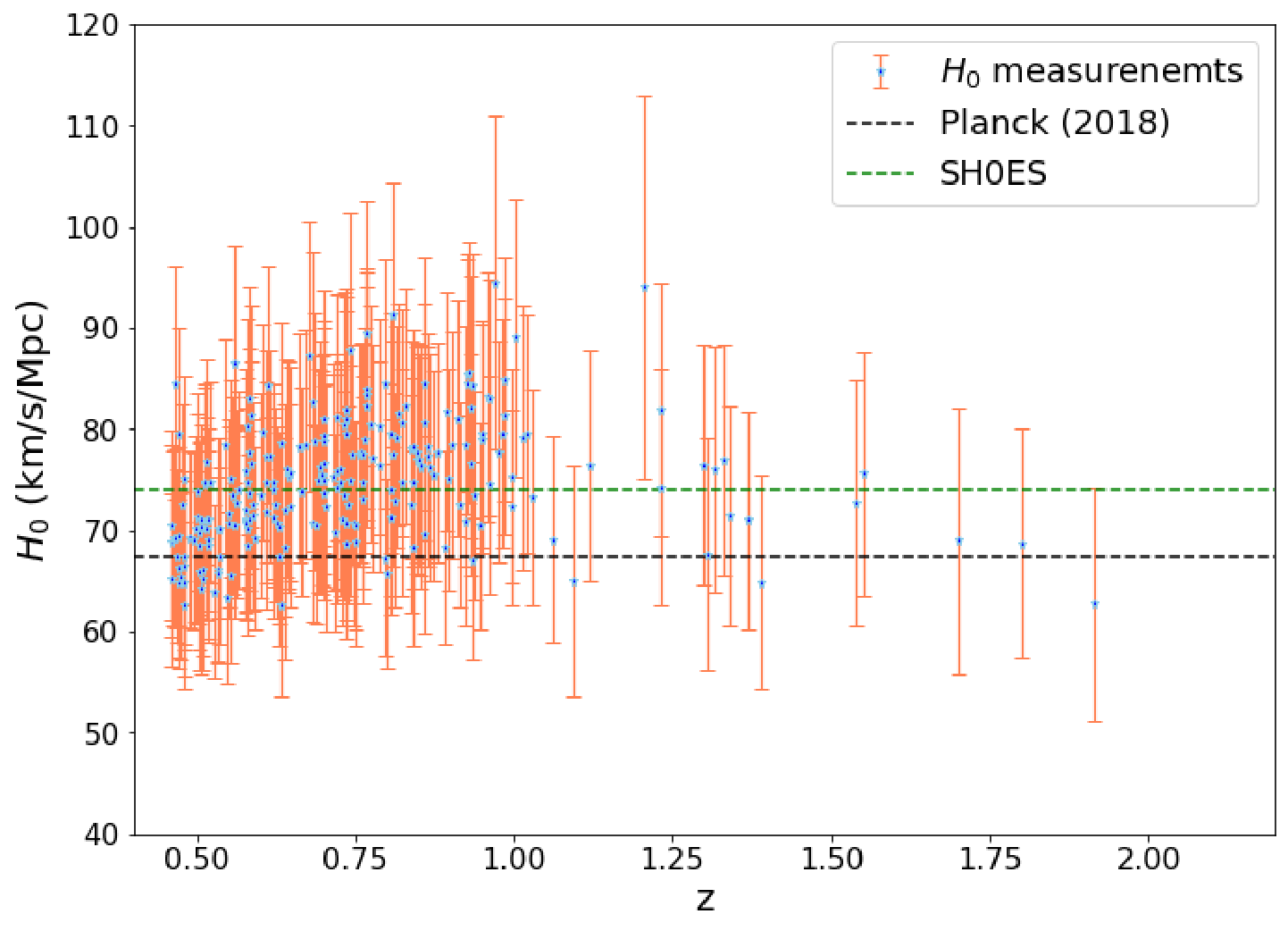}
\includegraphics[width=8cm,height=6.2cm]{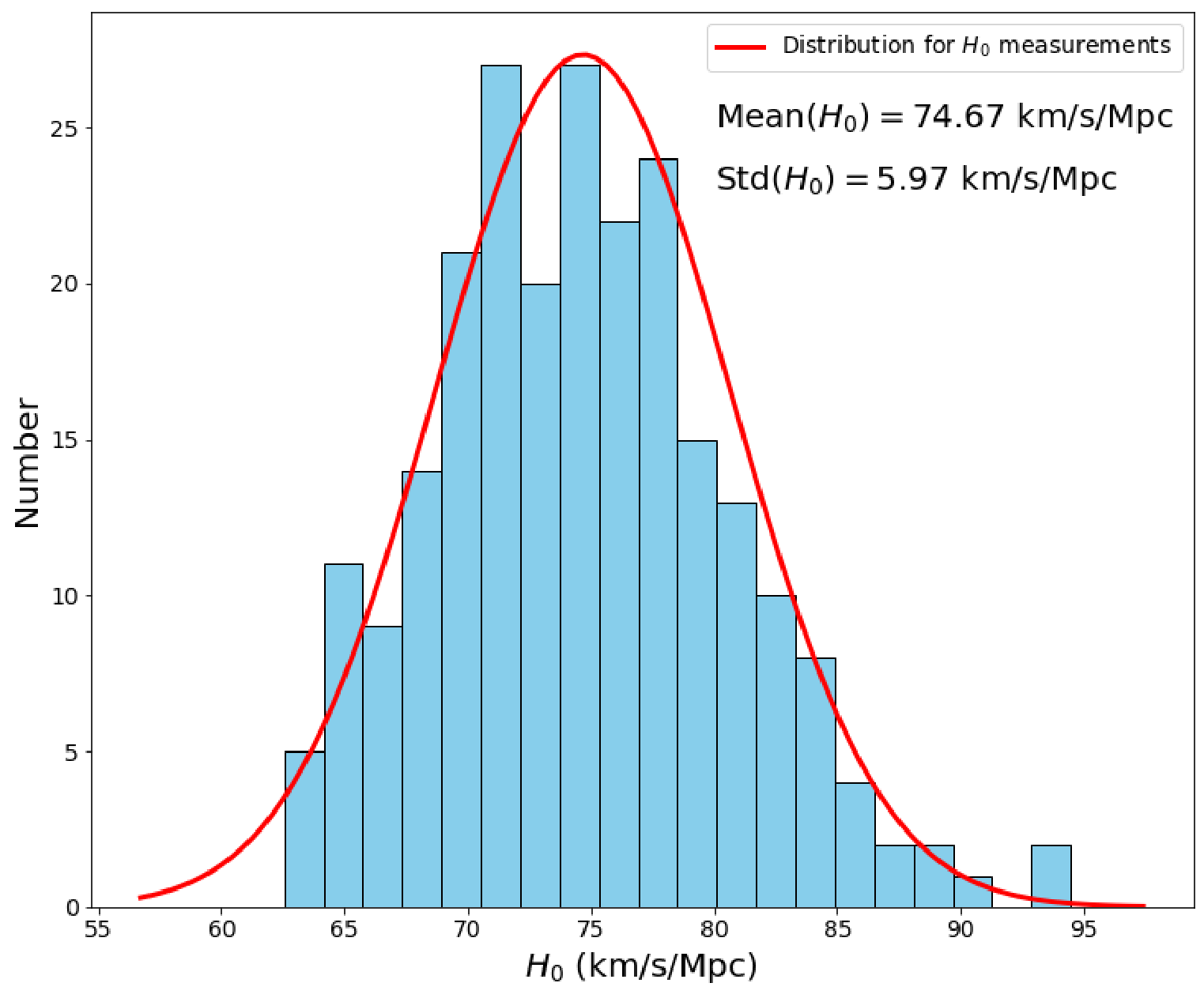}}
\caption{\textit{Left panel}: Individual measurements of $H_0$ based on the 237 SN Ia data-points with $d_A(z)$ reconstructed from the full sample of 120 radio quasars using ANN method.  \textit{Right panel}:  The histogram plot for the measurements of $H_0$ with best fit Gaussian distribution overplotted.}
\end{figure}

The individual measurements of $H_0$ are shown in the left panel of the Fig. 3. Right panel shows the histogram of $H_0$ measured. We have also tested it for Gaussianity. Kolmogorov-Smirnov test has the p-value of $p=0.523$ and Lilliefors test yields $p=0.164$. In both cases one cannot reject Gaussianity assumption. The difference in tests between Kolmogorov-Smirnov  and Lilliefors and between 37 vs. 237 data-points can be understood in terms of the skewness noticeable on the histogram in Fig.3. However, the meaningful use of the weighted mean summary is supported.
The results of different statistical approaches adopted by us are the following. The weighted mean and corresponding uncertainty are $\widetilde{H}_0=73.51(\pm0.67) \Mpc$.
By using the median value and corresponding median absolute deviation, we obtain the $Med(H_0)=74.71(\pm4.08) \Mpc$. This time both weighted mean and the median are consistent with each other. Moreover, the MAD region around the median differs by more than $5\sigma$ from the \emph{Planck} value ($\sigma$ referring to the \emph{Planck} result).
With the sample size increased substantially after using ANN reconstructed angular diameter distances, our result is totally consistent with the $H_0$ values inferred from local (i.e. not early Universe) probes, as reported by SH0ES and H0LiCOW collaborations.
MCMC assessment of $H_0$ posterior yields $H_0=73.52^{+0.66}_{-0.68} \Mpc$.
It is worth stressing that the technique we proposed, that is the use of data driven reconstruction of angular diameter distances via ANN achieved a of $1\%$, which is comparable to the result from \emph{Planck} 2018 TT, TE, EE+lowE+lensing data \citep{Aghanim2020}.  This is one of the most important conclusions of our work.

\begin{figure}
\centering
{\includegraphics[width=8cm,height=6.2cm]{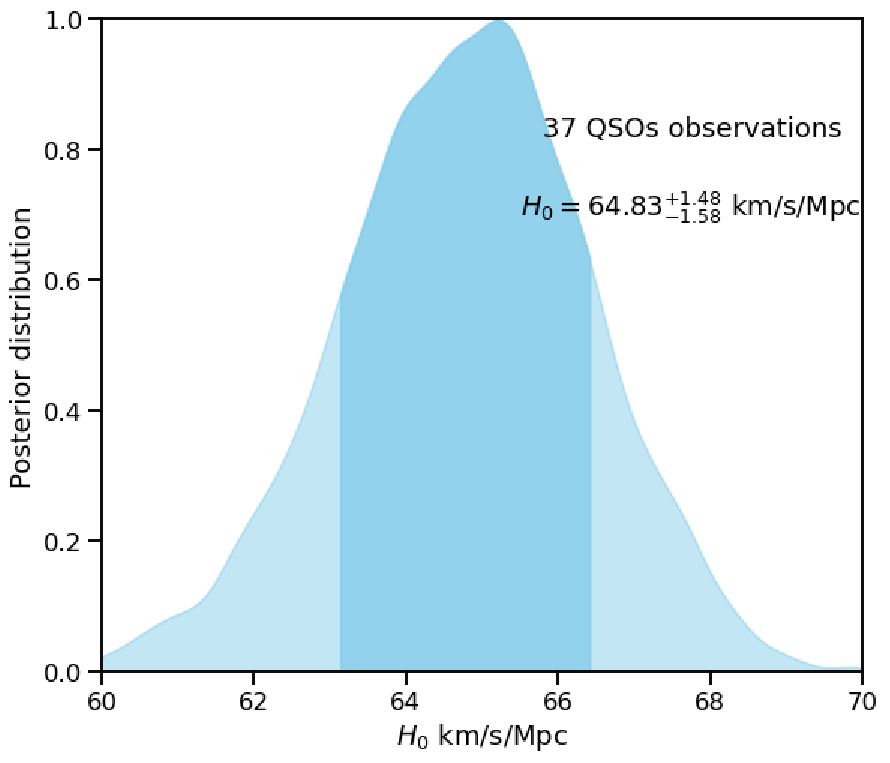}
\includegraphics[width=8cm,height=6.2cm]{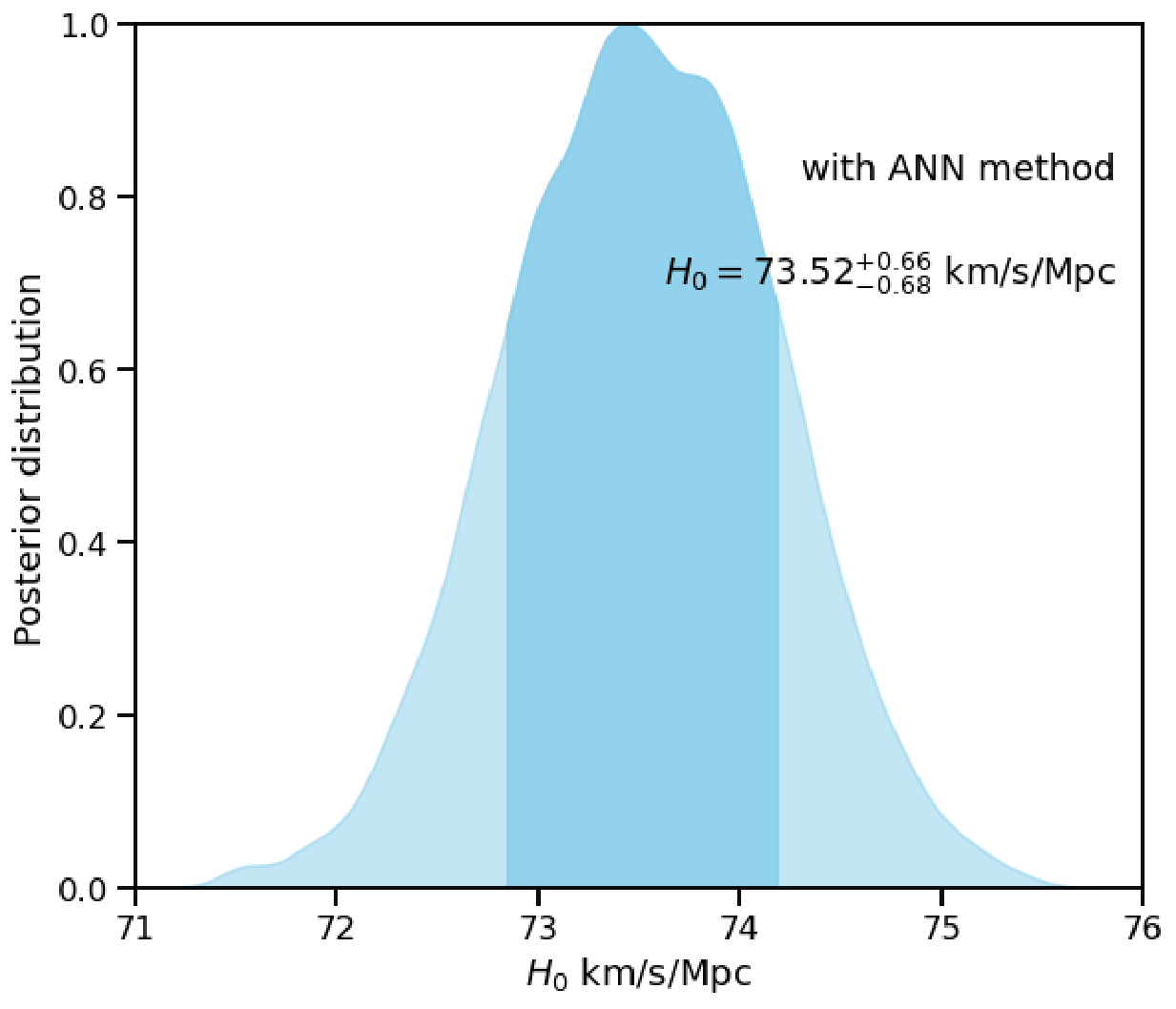}}
\caption{\textit{Left panel}: Posterior probability density of $H_0$ from MCMC simulations based on 37  radio quasar -- SN Ia pairs matched by redshift. \textit{Right panel}:  Posterior probability density of $H_0$ from MCMC simulations based on 237 SN Ia with $d_A(z)$ reconstructed using ANN method.}
\end{figure}

It is necessary to emphasize that our assessment of $H_0$ depends on the calibration of the linear size parameter $l$ in radio quasars and $a_B$ in SN Ia. Because the physical meaning of the compact structure size in radio quasars and the absolute B-band magnitude of SN Ia (whose value is determined by the host stellar mass) is not very clear, it is hard to determine the linear size parameter $l$ and SN Ia parameter $a_B$ precisely. Moreover, calibration of $l$ and $a_B$ by other astronomical probes would also introduce additional systematic uncertainties.
We have checked the influence of calibration parameters on our conclusions. Namely, the Table 1 of \cite{2017JCAP...02..012C} presents calibration parameters for different samples of radio-quasars using two cosmology independent methods. Regarding our sample of 120 intermediate-luminosity radio quasars the linear size parameter most different from the value we used is $l=10.86 \pm 1.58$. We check that adopting this value changes central values of $H_0$ by about $1.63\%$.
Similarly, if we use different value of SN Ia nuisance parameter $a_B=0.71719$ as in \cite{2016ApJ...826...56R},  it will affect the measurement accuracy of $H_0$ by about $1.12\%$.
The precision of $H_0$ assessment is not noticeably affected. Hence, our method offers a new way to measure $H_0$ with high precision.

Recently, a lot of attention has been paid on whether $H_0$ determined from direct observations of local probes up to high redshifts displays any trace of evolution with redshift \citep{2022Galax..10...24D,2020PhRvD.102j3525K,2021PhRvD.103j3509K,2022arXiv221200238J}. Let us stress that ``evolution'' of $H_0$ does not have sense: by definition $H_0$ is the present expansion rate, hence has one exact value. What we mean here is a way to discover possible systematic effects arising in using local probes at different redshifts. Any such effect, if significant evidence of it is found, most probably would indicate that calibrations on nearby objects is biased when applied to high redshift ones, e.g. metallicity effects on Cepheid calibration.
Hence, the term of ``evolving $H_0$'' should be treated as a useful mental shortcut.

For example, \citet{Dainotti21} used the Pantheon sample split into appropriate redshift bins to fit the extracted $H_0$ values with a function mimicking the redshift evolution $H_0(z)=H_0/(1+z)^{\alpha}$. They found that $H_0$ evolves with redshift, showing a slowly decreasing trend, with $\alpha$ coefficients consistent with zero only within $1.2 - 2.0\sigma$ confidence level. However, their  assessment of $H_0$ assumed a cosmological model as $\LCDM$ and $\omega_0\omega_a$CDM. Inspired by these findings, we decide to check whether our data, i.e. combined SN Ia and intermediate luminosity radio quasars support these claims. For this purpose, we take advantage of already mentioned circumstance that individual $H_0$ values are obtained at different redshifts spanning a considerable range. Moreover, the advantage is that we do not need to specify the cosmological model. Our approach is purely data driven.  We propose to test the redshift dependence of $H_0$ by considering the ratio: $\eta_{H_0} = H_{0,z_i}/H_{0,z_{fid}}$ where the $H_{0,z_{fid}}$ is some fiducial value. From Eq.(\ref{H0}) and Eq.(\ref{H0_new}) one can see that calibration parameters $a_B$ and $l$ cancel and do not enter directly to the test. Hence they do not introduce any additional systematics to determination of $\eta_{H_0}$. It is also important to note that whatever fiducial value we chose possible presence or lack of redshift dependence of $H_0$ should not be affected. In other words statistical consistency of our ratio with the value of $\eta_{H_0}=1$ would indicate lack of systematic redshift dependence of measured $H_0$.

In our analysis, we chose the point with the smallest redshift as the fiducial value.
The individual values of $\eta_{H_0}$ are shown in the Fig. 5. Already by inspection one can see that our results are consistent with $\eta_{H_0}=1$. The weighted mean value and corresponding uncertainty is $\eta_{\widetilde{H}_0}=1.05\pm0.05$. By using the median value and corresponding median absolute deviation, we obtain the $Med(\eta_{H_0})=1.07(\pm0.08) \Mpc$. Meanwhile, we also perform the same test choosing the median as a fiducial value. The final results are consistent with $\eta_{H_0}=1$ within corresponding uncertainty. Thus, adopting the median as the fiducial value do not change our main conclusion, therefore we do not show these results but conclude that the result we obtained is robust.

\begin{figure}
\centering
{\includegraphics[width=10cm,height=8cm]{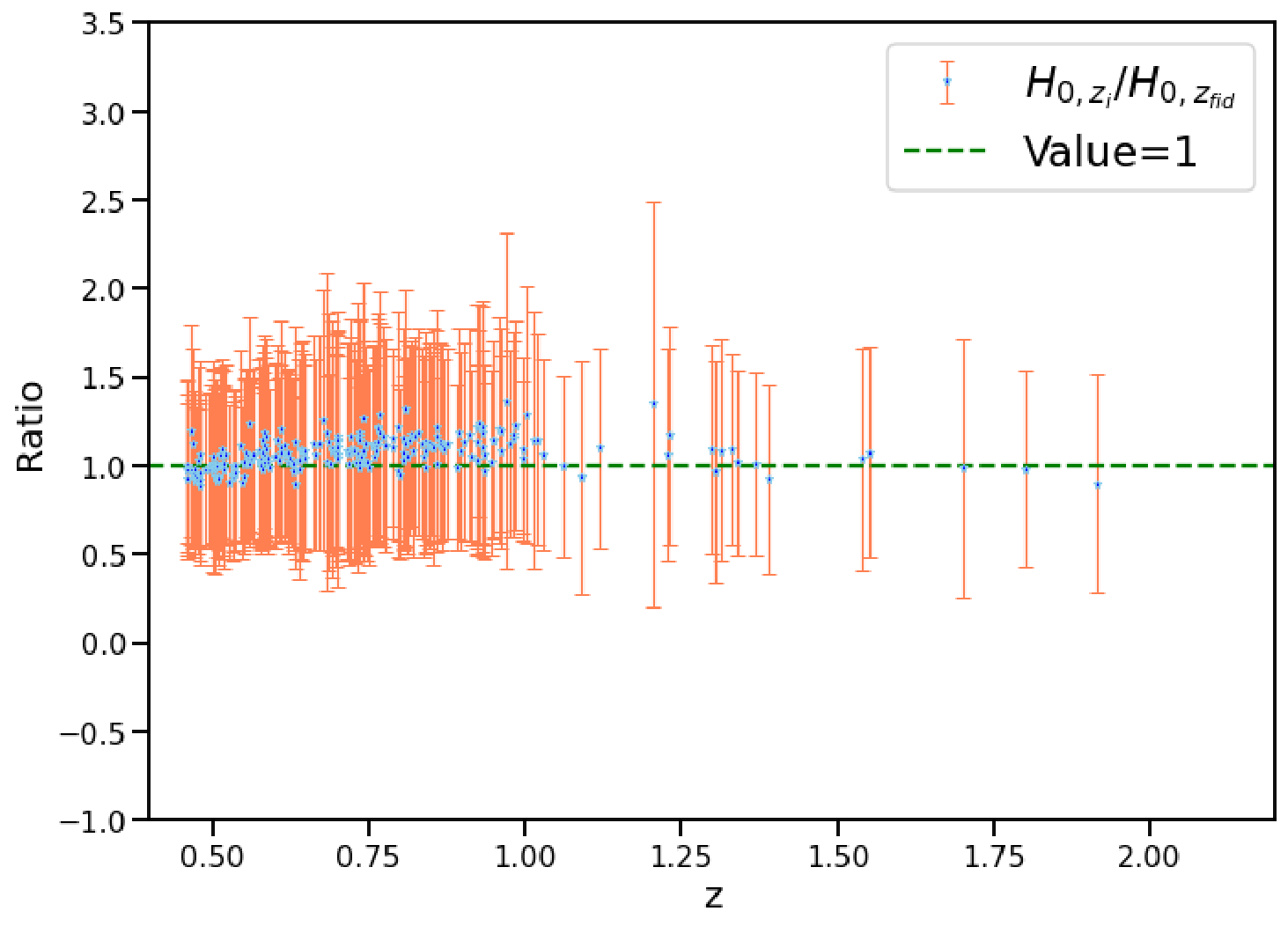}}
\caption{Scatter plot of the $\eta_{H_0}$ ratio based on the 237 observations of SN Ia with $d_A(z)$ reconstructed from radio quasars by ANN method.}
\end{figure}

\section{Conclusion}
In this letter, we use the data-sets of SN Ia (Pantheon) acting as standard candles together intermediate luminosity radio quasars acting as standard rulers to determine the Hubble constant according to the idea proposed by \citet{2023PhRvD.107b3520R}.
The idea is based on the distance duality relation between luminosity distance $d_L(z)$ and angular diameter distance $d_A(z)$, which is robust and independent of any cosmological model.
For this purpose, we recover the $d_A(z)$ relation using ANN trained on 120 radio quasar sample. In order to cover the same redshift range by SN Ia, the sub-sample of 237 supernovae is selected. The result is unambiguously consistent with values of $H_0$ obtained from local probes by SH0ES and H0LiCOW collaborations. Three statistical summary measures: weighted mean, median and MCMC simulated posterior distribution are fully consistent with each other and the precision reached $1\%$ level. This is encouraging for the future applications of our method. For the sake of illustration, we also considered 37 matched pairs of quasars and supernovae, where $d_A(z)$ was obtained directly, not from ANN reconstruction. In this case, weighted mean and posterior distributions are consistent with \emph{Planck} results. However, the sample is scatter dominated with large individual uncertainties. The most robust summary statistics i.e. the median and MAD turn out consistent with both local and \emph{Planck} values, although the median itself is consistent with SH0ES results. Because individual measurements of $H_0$ are related to different redshifts spanning the range from $z=0.5$ to almost $z=2.0$, we take advantage of this fact and our cosmological model independent method to check if there is any noticeable trend in $H_0$ measurements with redshift of objects used for this purpose. Our result is that the data alone strongly support the lack of such systematic effects.

As a final remark, we also look forward to a large amount of future data, not only from the radio quasars, but also from the SN Ia, allowing us to further improve the precision of $H_0$ measurements.  In the future, multi-frequency VLBI observations will yield more high-quality quasar observations based on better UV coverage \citep{2015MNRAS.452.4274P}. These radio quasars have more compact structures, higher angular resolution, and less statistical and systematic uncertainty. On the other hand, the Nancy Grace Roman Space Telescope is the future NASA mission. In a baseline 6-yr mission, including a 2-yr supernova survey strategy, it is expected to discover about $10^3\sim10^4$ SN Ia \citep{2018ApJ...867...23H}.  Future observations will create a perfect opportunity to apply the method presented here on much larger and better samples.
Meanwhile, considering the variety of different machine learning algorithms and fast progress in this area, we may also be optimistic in measuring $H_0$ with much higher precision.

\section*{Acknowledgments}
The authors are grateful to the referee for constructive comments, which allowed to improve the paper substantially.
Liu. T.-H was supported by National Natural Science Foundation of China under Grant No. 12203009;  Chutian Scholars Program in Hubei Province. Hubei Province Foreign Expert Project (2023DJC040);  Wang. J. C was supported by the National Natural Science Foundation of China under Grant No. 12122504 and No. 12035005.

\bibliographystyle{unsrt}
%\bibliography{Reference}

\end{document}